\documentclass[a4paper,11pt]{article}

\usepackage{CoDaWork}
\usepackage{bm}
\usepackage{epstopdf}

\title{ A data-based power transformation for compositional data}
\authors{{\bf Michail T. Tsagris,  Simon Preston and Andrew T.A. Wood} }
\normalsize 
\affiliation{Division of Statistics, School of Mathematical Sciences, University of Nottingham, UK; \email{pmxmt1@nottingham.ac.uk} }

\begin{document}
\maketitle

\section{Introduction}

A compositional data vector is a special type of multivariate observation in which the elements of the vector are non-negative and sum to a constant, usually taken to be unity. Data of this type arise in many fields including geology, archaeology, biology and economics.
In mathematical terms, the relevant sample space is the standard simplex, defined by 
\begin{eqnarray} \label{simplex}
S^d=\left\lbrace(x_1,...,x_D) \big| x_i \geq 0,\sum_{i=1}^Dx_i=1\right\rbrace  \ \ \text{where} \ \ d=D-1.
\end{eqnarray} 

The following two approaches have been widely used for compositional data analysis. The first is to neglect the compositional constraint and apply standard multivariate data analysis techniques to the raw data, an approach we will call raw data analysis (RDA). For discussion and examples of RDA see for example \citet{baxter1995, baxter2001, baxter2005, baxter2006}. RDA is also adopted by \citep{bio2007} in biological settings. The second approach, introduced by \citet{ait1982, ait1983, ait1986}, is based upon transforming the data using log-ratios, an approach we will call log-ratio analysis (LRA). The motivation behind LRA is that compositional data carry only relative information about the components and hence working with logs of ratios is appropriate for analysing this information. A popular log-ratio transformation, suggested by \citet{ait1983, ait1986}, is the centred log-ratio transformation defined by
\begin{eqnarray} \label{clr}
\mathbf{y}=\{y_i\}_{i=1,\ldots , D}=\left\lbrace\log \frac{x_i}{g({\bf x})}\right\rbrace_{i=1,\ldots , D}
\end{eqnarray}
where $g({\bf x})=\left(\prod_{j=1}^Dx_j\right)^{1/D}$ is the geometric mean of the $D$ components of the composition. This transformation maps the standard simplex (\ref{simplex}) onto a $d$-dimensional subspace of $R^D$, given by
\begin{eqnarray*}
Q^d=\left\lbrace \left(x_1,...,x_D\right) \big| \sum_{i=1}^Dx_i=0 \right\rbrace
\end{eqnarray*} 
The so called isometric log-ratio transformation \citep{ilr2003} is then given by ${\bf z} = {\bf Hy}$, where ${\bf H}$ is a $d \times D$ orthonormal matrix whose rows are orthogonal to ${\bf 1}_D$, the $D$-vector of ones. A standard choice of ${\bf H}$ is the Helmert sub-matrix obtained by removing the first row from the Helmert matrix \citep{helm1965}. 

Over the years there has been a debate over whether it is preferable to use RDA or LRA \citep{barcelo1996, ait1999, ait2000, beardah2003, sharp2006}. In this paper we take the view that the choice between RDA, LRA and other possibilities should depend at least in part on the data, and should not be purely a matter of \textit{a priori} considerations. For this purpose we consider a framework in which RDA and LRA relate to two special cases of a one-parameter family of transformations of the simplex (\ref{simplex}). The idea is that we should explore this family of transformations when deciding which approach to use for a given compositional dataset because, for different datasets, different transformations may be preferable. For a more extensive discussion of this framework we refer to \citet{mike2011}. 
  
The outline of this paper is as follows. Section 2 contains a short and informal comparison of RDA and LRA, while in section 3 we see the contrasting results of applying RDA and LRA to some examples involving real and artificial data. In section 4, we describe this family of power transformations, discuss its relationship with RDA and LRA and show an example of analysis based on these transformations. Finally, in section 5, we present our conclusions.

\section{An informal comparison of RDA and LRA}

\subsection{Simplicial distance}
\citet{ait1992} argues that a simplicial metric should satisfy certain properties. These properties include 

\begin{enumerate}
\item \textit{Scale invariance}. The requirement here is that the measure used to define the distance between two compositional data vectors should be scale invariant, in the sense that it makes no difference whether, for example, the compositions are represented by proportions or percentages.
\item \textit{Subcompositional dominance}. To explain this we consider two compositional data vectors, and select subvectors from each consisting of the same components. Then subcompositional dominance means that the distance between the subvectors is always less than or equal to the distance between the original compositional vectors.
\item \textit{Perturbation invariance}. The perturbation of one compositional data vector by another is defined by (\ref{perturbation}) below. The requirement here is that the distance between compositional vectors ${\bf x}$ and ${\bf w}$ should be the same as distance between ${\bf x}\oplus_0{\bf p}$ and ${\bf w}\oplus_0{\bf p}$, where the operator $\oplus_0$ is defined in (\ref{perturbation}) and ${\bf p}$ is any vector with positive components.
\end{enumerate}

RDA amounts to using the Euclidean distance as a metric in the simplex. In RDA the distance between two vectors ${\bf x}$ and ${\bf w} \in S^d$ is 
\begin{eqnarray} \label{rawdist}
\Delta_{RDA}({\bf x, w })=\left[\sum_{i=1}^D\left(x_i-w_i\right)^2\right]^{1/2}.
\end{eqnarray} 

In LRA the relevant metric is the Euclidean distance applied to the log-ratio transformed data \citep{ait1983, ait1986}. In mathematical terms the expression is:
\begin{eqnarray} \label{clrdist}
\Delta_{LRA}({\bf x, w })=\left[\sum_{i=1}^D\left(\log\frac{x_i}{g({\bf x})}-\log\frac{w_i}{g({\bf w})}\right)^2\right]^{1/2}.
\end{eqnarray}

\citet{ait1986} has shown that $\Delta_{LRA}$ satisfies the three properties above. It is straightforward to see that, in contrast to $\Delta_{LRA}$, $\Delta_{RDA}$ does not satisfy any of these properties (although $\Delta_{RDA}$ is equivariant with respect to a scale change, in an obvious sense). On this basis, Aitchison argued that the metric $\Delta_{LRA}$ is preferable to $\Delta_{RDA}$. However, although at first glance properties 1-3 seem to be reasonable requirements, there will sometimes be a price to be paid for using $\Delta_{LRA}$. In particular, this choice and similarly the choice of $\Delta_{RDA}$, ties us to a specific geometry on the simplex which may or may not be appropriate for a given dataset. Our thinking here is that, ideally, one should work with a geometry on the simplex in which the structure of the data is "as nice as possible". In the data examples considered in section 3, niceness means linear structure and a measure of central tendency which lies in the main body of the sample, though of course different definitions of niceness could be adopted. The main point, however, is that one does not know in advance what the most appropriate geometry on the simplex is, because it will depend on the data.

\subsection{Measures of central tendency and simplicial addition}
The two approaches lead to different definitions of measures of central tendency. The means specified in (\ref{mean}) and (\ref{geo}) below are Fr{\'e}chet means with respect to $\Delta_{RDA}$ and $\Delta_{LRA}$, respectively. The Fr{\'e}chet mean on a metric space $\left(\mathcal{M}, \text{dist} \right)$, with the distance between $p,q\in \mathcal{M}$ given by $\text{dist}\left(p,q\right)$, is defined by   
$\text{argmin}_hE_X\left[\text{dist}\left({\bf X}, h\right)^2 \right]$ and $\text{argmin}_h\sum_{i=1}^n\text{dist}\left(x_i, h\right)^2$
in the population and finite sample cases, respectively. In the geometry determined by RDA, we define the simplicial analogue of addition, denoted by $\oplus_1$, as vector addition followed by rescaling to fix the sum of components to be 1.  

\begin{eqnarray*}
{\bf x}\oplus_1{\bf w}=\mathcal{C}\left\lbrace\left\lbrace x_i + w_i\right\rbrace_{i=1,\ldots , D}\right\rbrace
\end{eqnarray*}
where $\mathcal{C}$ is the closure operation $\mathcal{C}\left({\bf x}\right)={\bf x}/\left(x_1+ \ldots +x_D \right)$. The mean in this case is defined as the simple arithmetic mean of the components: 
\begin{eqnarray} \label{mean} 
\bm{\mu}_{\left(1\right)}=\left\lbrace\frac{1}{n}\sum_{j=1}^nx_{ij}\right\rbrace_{i=1,\ldots , D}. 
\end{eqnarray}

In the geometry defined by LRA, the analogue of addition of two compositional vectors ${\bf x}$ and ${\bf w}$, known as perturbation of ${\bf x}$ and ${\bf w}$ which we denote by ${\bf x}\oplus_0{\bf w}$, is given by the component-wise product followed by the closure operation:  
\begin{eqnarray} \label{perturbation}
{\bf x}\oplus_0{\bf w}=\mathcal{C}\left\lbrace\left\lbrace x_iw_i\right\rbrace_{i=1,\ldots , D}\right\rbrace.
\end{eqnarray} 
In the LRA case, the Fr{\'e}chet mean is given by the vector consisting of the sample geometric mean of each component followed by the closure operation \citep{ait1989}: 
\begin{eqnarray} \label{geo}
\bm{\mu}_{\left(0\right)}=\mathcal{C}\left\lbrace\left\lbrace\prod_{j=1}^nx_{ij}^{1/n}\right\rbrace_{i=1,\ldots , D}\right\rbrace.
\end{eqnarray} 

\subsection{Further discussion}
LRA works very well when the data follow the logistic normal distribution; see for instance Hongite, Kongite and Boxite data analysed by \citet{ait1986}. However it is not necessarily clear whether for data with different distributions that the LRA is still the best approach. \citet{baxter2006} performed principal component analysis on both simulated and real data. The conclusion they reached was, in the examples considered, the RDA produced more archaeologically interpretable results than LRA. In a previous paper \citet{beardah2003} mention that the use of PCA on standardized data may be regarded as incorrect by proponents of LRA but did recover interpretable structure whereas the use of the log-ratio transformation failed to detect it. The standardization of the data after the centred log-ratio transformation worked well but the produced results are similar to the standardized data analysis for the most part. 

RDA has another potential advantage: in contrast to the LRA mean in (\ref{geo}), the RDA mean (\ref{mean}) is well-defined even when the data have some components equal to zero.  

A general transformation applied to compositional data is the Box-Cox transformation applied to ratios of components
\begin{eqnarray} \label{BC}
\mathbf{v}=\{v_i\}_{i=1,\ldots , D}=\left\lbrace\frac{\left(\frac{x_i}{x_D}\right)^{\lambda}-1}{\lambda}\right\rbrace_{i=1,\ldots , D}. 
\end{eqnarray}
The $x_D$ stands for the last component but with relabelling any component can play the role of the common divisor. As $\lambda \rightarrow 0$, the above expression tends to the "additive log-ratio transformation" which was defined by \citet{ait1986} as an alternative transformation to the centred log-ratio transformation in (\ref{clr})
\begin{eqnarray} \label{alr}
\mathbf{v}=\{v_i\}_{i=1,\ldots , D}=\left\lbrace\log \frac{x_i}{x_D}\right\rbrace_{i=1,\ldots , D}. 
\end{eqnarray}
In \citet{barcelo1996} the Box-Cox transformation defined in (\ref{BC}) is examined and the conclusion was that some samples cannot be adequately modelled using the additive log-ratio transformation defined in (\ref{alr}), in the sense that the value of $\lambda$ should not be restricted to zero. 

Recently \citet{sharp2006} suggested the graph median a measure of central tendency alternative to the geometric or the arithmetic mean. His purpose was the proposal of a suitable alternative to the closed geometric mean when the latter fails in the sense that it does not lie within the body of the data.    

\section{Some examples}

We now present some examples with real and artificial data. In the first example the use of LRA is to be preferred whereas RDA seems to work better in the second example.  
 
\subsection{Example 1: Hongite data}

\begin{figure}[!h]
\begin{center}
\includegraphics[width=0.5\textwidth]{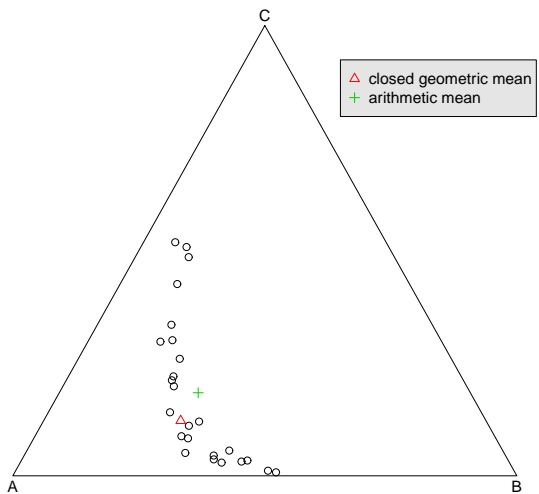}
\end{center}
\caption{Ternary diagram of a subcomposition of the Hongite data. The closed geometric mean (\ref{geo}) and the arithmetic mean (\ref{mean}) are presented.}
\label{fig1}
\end{figure}

We created a subcomposition taking the first 3 components of Hongite data \citet{ait1986} and plotted it in Figure \ref{fig1}. The closed geometric mean in (\ref{geo}) clearly lies within the body of the data, whereas the arithmetic mean in (\ref{mean}) has failed in this example as it lies outside the body. The assumption of logistic normality based upon the battery of tests suggested by \citet{ait1986} is not rejected for this composition.    
 
\newpage
\subsection{Example 2: Artificial data}

\begin{figure}[!h]
\begin{center}
\includegraphics[width=0.5\textwidth]{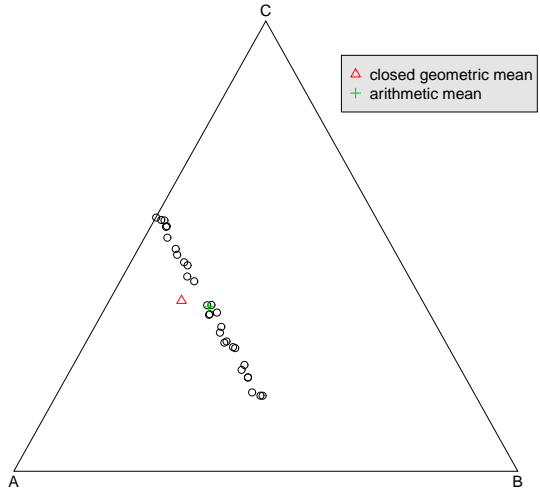}
\end{center}
\caption{Ternary diagram of artificial data showing an example where the geometric mean appears to be an unsuitable measure of central tendency.}
\label{fig2}
\end{figure}

Here, we created a dataset in which is concentrated on a straight line when plotted in a ternary diagram; see Figure \ref{fig2}. In this case, the data do not follow the logistic normal distribution based upon the battery of tests suggested by \citet{ait1986}. The arithmetic mean (\ref{mean}) is a good measure of central tendency, whereas the closed geometric mean (\ref{geo}) has failed because it lies off the line around which the data are concentrated.  

Our conclusions from the examples considered above are that  in the first case, the approach based on LRA and the closed geometric mean seems more appropriate, while in the second case RDA and the arithmetic mean seems better. In general, the question of interest is to decide whether to use LRA, RDA or perhaps another possibility. We believe that this decision should depend on the nature of the data at hand.

\section{A family of power transformations on the simplex}

We now consider a Box-Cox type transformation in terms of which the RDA and LRA can be considered as special cases. Consider the power transformation introduced by \citet{ait1986}
\begin{eqnarray} \label{alpha}
{\bf u}=\{u_i\}_{i=1, \ldots , D}=\left\lbrace\frac{x_i^{\alpha}}{\sum_{j=1}^Dx_j^{\alpha}}\right\rbrace_{i=1, \ldots, D},
\end{eqnarray}
\\
and in terms of this define: 
\begin{eqnarray} \label{isoalpha}
{\bf z}=\frac{1}{\alpha}\left(D{\bf u}-{\bf 1}\right){\bf H}^T, 
\end{eqnarray} 
where we take ${\bf H}$ to be the $d \times D$ Helmert sub-matrix mentioned in the Introduction. Note that the $\alpha$-transformed vector ${\bf u}$ in (\ref{alpha}) remains in the simplex $S^d$, whereas ${\bf z}$ is mapped onto a subset of $R^d$. Note that (\ref{isoalpha}) is simply a linear transformation of (\ref{alpha}). Moreover as $\alpha \rightarrow 0$, (\ref{isoalpha}) converges to the isometric log-ratio transformation. For a given $\alpha$ we can define a simplicial distance:   
\begin{eqnarray} \label{adist}
\Delta_{\alpha}({\bf x, w})=\frac{D}{\alpha}\left[\sum_{i=1}^D\left(\frac{x_i^\alpha}{\sum_{j=1}^D x_j^\alpha}-\frac{w_i^\alpha}{\sum_{j=1}^D w_j^\alpha}\right)^2\right]^{1/2}.
\end{eqnarray} 
Associated with this one-parameter family of distances is the family of Fr{\'e}chet means

\begin{eqnarray} \label{frechet}
\bm{\mu}_{\left(\alpha\right)}=\mathcal{C}\left\lbrace\left\lbrace\left(\frac{1}{n}\sum_{j=1}^n\frac{x_{ij}^\alpha}{\sum_{k=1}^Dx_{kj}^\alpha}\right)^{1/\alpha}\right\rbrace_{i=1,\ldots , D}\right\rbrace.
\end{eqnarray}
This agrees with (\ref{mean}) when $\alpha=1$ and it agrees with (\ref{geo}) when $\alpha=0$. Specifically, as $\alpha \rightarrow 0$, the $\alpha$-distance (\ref{adist}) and Fr{\'e}chet mean (\ref{frechet}) converge to Aitchison's distance (\ref{clrdist}) and the closed geometric mean (\ref{geo}), respectively. When $\alpha=1$, the Fr{\'e}chet mean is equal to the arithmetic mean of the components (\ref{mean}) and the $\alpha$-distance is equal to the Euclidean distance in the simplex (\ref{rawdist}) multiplied by $D$. 

The choice of the parameter $\alpha$ should depend on the type of analysis, we wish to perform. In discriminant analysis, for instance, the percentage of correct classification (estimated via cross-validation) could be used as a criterion to maximize with respect to the value of $\alpha$. Similarly in linear regression the pseudo-$R^2$ \citep{ait1986} could serve as a criterion.  

\begin{figure}[!h]
\begin{center}
 \includegraphics[width=0.5\textwidth]{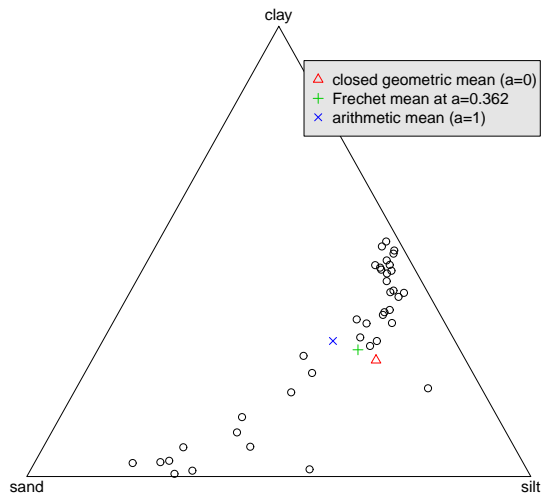}
\end{center}
 \caption{Ternary diagram of the Arctic lake data. The three Fr{\'e}chet means are plotted.}
 \label{fig3}
\end{figure}

Alternatively, if we are prepared to assume that the data follow a parametric model after an (unknown) $\alpha$-transformation, we could use the profile log-likelihood to choose $\alpha$. As in the Box-Cox transformation, the value of the parameter is selected by maximizing the profile log-likelihood of the transformed data. One possibility, which we adopted with the Arctic lake data \citep{ait1986} in Figure \ref{fig3}, was to treat the data as being from a (singular) multivariate normal after a suitable $\alpha$-transformation. This approach has the obvious drawback that it ignores the fact that any multivariate normal will assign positive probability outside the simplex, which may or not be of practical importance, depending on how concentrated the normal distribution is. The profile log-likelihood of $\alpha$ applied to (\ref{isoalpha}) is the same as the log-likelihood applied to (\ref{alpha}) plus a constant equal to $\frac{n}{2}\log{D}$, where $n$ and $D$ denote the sample size and the number of components of the composition respectively. Figure \ref{fig3} shows the ternary diagram of the Arctic lake data \citep{ait1986}. The profile log-likelihood of $\alpha$ for this data was maximized when $\alpha=0.362$. Thus the three Fr{\'e}chet means for $\alpha=0, 0.362 \ \text{and} \ 1$ were calculated and plotted as well. 

\newpage
\section{Conclusions}
Two standard approaches to compositional data analysis, referred to in the paper as RDA and LRA, were discussed and compared. On the basis of simple numerical examples and further considerations we concluded that which approach, if either, is to be preferred should depend on the data under study. We then considered a framework involving a one-parameter family of Box-Cox type power transformations which includes the RDA and LRA approaches as special cases. Depending on the purposes of the analysis, we may then choose an optimal value of $\alpha$ by optimizing a suitable criterion, and/or explore the practical implications of different choices of $\alpha$. This framework offers greater flexibility than simply adopting one method or another.  

\newpage
\section*{Appendix}
\begin{table}[!h]
\centering
\caption{Artificial data used in example 2}
\begin{tabular}{c c c c}
\hline
     &  Component A &   Component B   &   Component C  \\ \hline
A1   &  0.4355095   &   0.3392920924  &	 0.2251984   \\
A2   &  0.4388882   &   0.2135973967  &	 0.3475144   \\
A3   &  0.4266460   &   0.0305602623  &	 0.5427937   \\
A4   &  0.4225122   &   0.4097265913  &	 0.1677612   \\
A5   &  0.4240518   &   0.3024372049  &	 0.2735110   \\
A6   &  0.4315424   &   0.3605322539  &  0.2079254   \\
A7   &  0.4337943   &   0.2779268820  &	 0.2882788   \\
A8   &  0.4358406   &   0.0006831432  &	 0.5634762   \\
A9   &  0.4300394   &   0.1057353415  &	 0.4642253   \\
A10  &  0.4279817   &   0.2512896242  &	 0.3207287   \\
A11  &  0.4310879   &   0.1469908804  &	 0.4219212   \\
A12  &  0.4206820   &   0.2268498575  &  0.3524682   \\
A13  &  0.4396177   &   0.3852283802  &	 0.1751539   \\
A14  &  0.4265380   &   0.1160721475  &	 0.4573898   \\
A15  &  0.4371975   &   0.2550170900  &	 0.3077854   \\
A16  &  0.4244092   &   0.0314046749  &	 0.5441861   \\
A17  &  0.4320087   &   0.1992476493  &	 0.3687437   \\
A18  &  0.4394902   &   0.2748886793  &	 0.2856211   \\
A19  &  0.4283819   &   0.0134845902  &	 0.5581335   \\
A20  &  0.4319368   &   0.0743081935  &	 0.4937550   \\
A21  &  0.4232230   &   0.2074545214  &	 0.3693225   \\
A22  &  0.4309724   &   0.3602299476  &	 0.2087976   \\
A23  &  0.4272409   &   0.2970880808  &	 0.2756710   \\
A24  &  0.4359625   &   0.0453307017  &  0.5187068   \\
A25  &  0.4267075   &   0.4053411348  &	 0.1679513   \\
A26  &  0.4396363   &   0.1278193992  &	 0.4325443   \\
A27  &  0.4355290   &   0.0837220492  &	 0.4807489   \\
A28  &  0.4226007   &   0.0204375902  &	 0.5569617   \\
A29  &  0.4366681   &   0.2144233973  &	 0.3489085   \\
A30  &  0.4242957   &   0.3395783944  &	 0.2361259   \\ \hline
Closed geometric mean  &  0.4778500  &  0.1432412430  &  0.3789087 \\
Arithmetic mean        &  0.4306997  &  0.2038899384  &  0.3654103 \\ \hline
\end{tabular}
\end{table}

\newpage
\bibliographystyle{chicago}   

\end{document}